\documentclass[12pt]{article}
\begin{document}

\begin{center}
{\large \bf An Information-Theoretic Link Between Spacetime Symmetries and Quantum Linearity}
\end{center}
\vspace{0.1in}

\begin{center}

{Rajesh R. Parwani\footnote{Email: parwani@nus.edu.sg}}

\vspace{0.3in}

{Department of Physics and\\}
{University Scholars Programme,\\}
{National University of Singapore,\\}
{Kent Ridge, Singapore.}

\vspace{0.3in}
\end{center}
\vspace{0.1in}
\begin{abstract}

A nonlinear generalisation of Schrodinger's equation is obtained using information-theoretic arguments.  The nonlinearities are controlled by an intrinsic length scale and involve derivatives to all orders thus making the equation mildly nonlocal. The nonlinear equation is homogeneous, separable, conserves probability, but is not invariant under spacetime symmetries. Spacetime symmetries are recovered when a dimensionless parameter is tuned to vanish, whereby linearity is simultaneously established and the length scale becomes hidden. It is thus suggested that if, in the search for a more basic foundation for Nature's Laws, an inference principle is given precedence over symmetry requirements, then the symmetries of spacetime and the linearity of quantum theory might both be emergent properties that are intrinsically linked. Supporting arguments are provided for this point of view and some testable phenomenological consequences highlighted. The generalised Klien-Gordon and Dirac equations are also studied, leading to the suggestion that nonlinear quantum dynamics with intrinsically broken spacetime symmetries might be relevant to understanding the problem of neutrino mass(lessness) and oscillations: Among other observations, this approach hints at the existence of a hidden discrete family symmetry in the Standard Model of particle physics.

\end{abstract}

\vspace{0.5in}

\section{Physics Through Inference}

A conceptually appealing and mathematically elegant method for deriving the form of probability distributions in statistical mechanics is Jaynes'  principle of maximum entropy \cite{jay}. The Gibbs-Shannon entropy (or information)
\begin{equation}
I_{GS} = -\int p(x) \ln p(x) \ dx \label{gs}
\end{equation}
is maximised under given constraints to determine the form for the probability distribution $p(x)$. For example, if the mean energy of the system $E = \int \epsilon (x) \ p(x) \ dx$ is specified, then introducing the Lagrange multiplier $\beta$ and maximising $I_{GS} - \beta E$ with respect to variations in $p(x)$ gives the well known canonical probability distribution $p(x) \propto \exp(-\beta \epsilon (x))$. In physics the quantity (\ref{gs}) is associated with the names of Bolzmann and Gibbs and applications are usually restricted to statistical mechanics. However the form (\ref{gs}) was derived independently by Claude Shannon \cite{Shannon} in his search for a measure that could be used to quantify the information content, or uncertainty, in a system 

It must be emphasized that the word ``entropy" used in relation to Shannon's information measure is an historical accident: It does not necessarily refer to the concept as used in statistical mechanics but rather to the more general idea of uncertainty. The principle of maximum entropy, or maximum uncertainty, is currently used as a method of inference in various fields of study, far beyond its original applications  \cite{max,Kapur}. 

If one already has some {\it a priori} information about the system in the form of a reference probability distribution $r(x)$, then the relative-entropy or Kullback-Leibler information \cite{Kullback}, 
\begin{equation}
I_{KL}(p,r) = -\int p(x) \ln {p(x) \over r(x)} \ dx \label{kl}
\end{equation}
is the relevant measure\footnote{The reader is alerted to differing sign-conventions and terminology in the literature.} that has to be maximised under the given constraints so as  to obtain explicit forms for $p(x)$. The negative of the Kullback-Liebler measure is also called the cross-entropy and is actually a distance measure, measuring the distance between the two probability distributions $p(x)$ and $r(x)$. If there is no useful {\it a priori} information then $r(x)$ can be taken to be a uniform distribution and the Kullback-Liebler measure then reduces to the Gibbs-Shannon entropy.

Thus the maximum entropy/uncertainty principle is a method of inference whereby one chooses probability distributions that provide the most unbiased description of the state of the system, since maximising the entropy measure acknowledges our ignorance of a more detailed structure. Since quantum mechanics also involves probabilities, it is natural to wonder if the Schrodinger equation, which describes the evolution of probability amplitudes, too can be derived using a method of inference. An attempt in this direction was made by Frieden \cite{fried1} but the full time-dependent Schrodinger equation was obtained only recently by Reginatto \cite{reg1}.

Consider the Schrodinger equation for a particle in one dimension,

\begin{equation}
i \hbar {\partial \psi \over \partial t} = - {{\hbar}^2 \over 2m} {\partial^2 \psi \over \partial x^2} + V(x) \psi \, . \label{sch1}
\end{equation}
The Madelung transformation \cite{mad} $\psi = \sqrt{p} \ e^{iS/ \hbar}$ can be used to write the Schrodinger equation in terms of two real functions, $p$ and $S$, called hydrodynamical variables,

\begin{eqnarray}
{\partial S \over \partial t} + {1 \over 2m} \left( {\partial S \over \partial x} \right)^2 + V + Q &=& 0 \, ,  \label{hj1} \\
{\partial p \over \partial t} + {1 \over m} {\partial \over \partial x} \left(p {\partial S \over \partial x} \right)&=&0 
\, ,  
 \label{cont1}
\end{eqnarray}
where

\begin{eqnarray}
Q &=&  - {{\hbar}^2 \over 2m}  {1 \over \sqrt{p}} {\partial^2 \sqrt{p} \over \partial x^2} \, \; \label{pot1} \\
&=&  - {{\hbar}^2 \over 8m} \left( {2 \partial^{2}_{x} p \over p} - { (\partial_{x} p  )^2 \over p^2} \right)
\end{eqnarray}
is called the quantum potential \cite{Ballentine, holland}. For $Q=0$, the  equation (\ref{hj1}) is just the Hamilton-Jacobi equation for a classical ensemble of particles described by a probability distribution $p(x,t)$ and with the function $S(x,t)$ related to the velocity of a particle by $v= {1 \over m} {\partial S \over \partial x}$. The equation (\ref{cont1}) is then just the continuity equation, equivalent to the expression for conservation of probability in quantum mechanics.

Reginatto notes that the classical limit of equations (\ref{hj1},\ref{cont1}), that is when $Q=0$, may be obtained by minimising the action 
\begin{equation}
\Phi_A \equiv \int p \left( {\partial S \over \partial t} + {1 \over 2m} \left( {\partial S \over \partial x} \right)^2 + V \right) \ dx \ dt \,
\end{equation}  
through a variation of both $p$ and $S$. The full quantum equations follow if the variation of the action $\Phi_A$ is restricted so that a particular measure of information, the Fisher information\footnote{This differs from (\ref{fish2}) below because in quantum mechanics the probability depends on the parameter $t$ in addition to the variable $x$.}, 
\begin{eqnarray}
I_F &=& \int {1 \over p} \left( {\partial p \over \partial x} \right)^2 \ dx \ dt \label{fish}
\end{eqnarray}
is also minimised. That is, minimising the linear combination $\Phi_A + \xi I_F $ with respect to both $p$ and $S$ and choosing the Lagrange multiplier $\xi = {\hbar}^2 / 8m$ gives precisely the equations (\ref{hj1},\ref{cont1})  which are equivalent to the time-dependent Schrodinger equation. It is crucial to note that the quantum potential (\ref{pot1}) arises solely from the variation of the information measure $I_F$. Thus one may argue that the method of inference allows one to supplement classical mechanics with the required fluctuations to arrive at quantum mechanics. The method works as well for many particles in higher dimensions \cite{reg1}.

The Fisher information measure \cite{fisher} is not commonly used in physics but is well-known in statistics where it is used in studies of parameter estimation \cite{Kullback} \footnote{The first few chapters of Ref.\cite{fried2} provide a concise introduction to Fisher information while the rest of that book develops a very different approach from that discussed in \cite{reg1} or this paper.}. Here is a brief description of the measure. Suppose one wishes to form an estimate of a parameter $\theta$ by making some measurements, $y$, in the presence of an additive intrinsic noise $x$, that is $y=\theta +x$. If $p(x)$ is the probability distribution of the noise, then the Cramer-Rao bound relates the variance of an unbiased estimate of $\theta$ to the Fisher information  ${\cal{I}_F}$,
\begin{equation}
\mbox{Var}(x) \ge {1 \over {\cal{I}}_F } \, 
\end{equation}   
with 
\begin{equation} 
{\cal{I}}_F = \int {1 \over p} \left( {\partial p \over \partial x} \right)^2 \ dx \, \label{fish2}.
\end{equation}
Minimising the Fisher information therefore implies choosing the probability distribution that maximises the mean-square error in the estimation of $\theta$, so that the distribution is one that is as unbiased as possible. Thus the principle of minimum Fisher information is similar in spirit to the principle of maximum entropy and Reginatto's derivation of Schrodinger's equation may be viewed as an extension of the method of inference to quantum mechanics.

As an aside, the approach of using the Hamilton-Jacobi equations and the Fisher information sheds light on some puzzling aspects of quantum mechanics. Firstly, since the real equations (\ref{hj1},\ref{cont1}) have a natural interpretation, it suggests that the use of {\it complex} numbers in quantum mechanics  is merely a convenience allowing one to combine the two real and nonlinear equations into one linear equation. Secondly, the form of the Fisher information measure suggests a reason \cite{fried2} as to why probability {\it amplitudes} play a fundamental role in quantum mechanics: The  Fisher information term $I_F$ becomes a polynomial function when written in terms of $p^{1/2}$,
\begin{equation}
I_F =  4 \int \left( {\partial p^{1 \over 2} \over \partial x} \right)^2 \ dx \ dt \, .
\end{equation}
Thirdly, the success of the approach in explicitly using ensembles and inference to derive the Schrodinger equation, suggests that the wavefunction $\psi$ does not represent reality but rather our limited knowledge of reality (the inference bit) and that it represents a virtual ensemble rather than a single particle (the ensemble bit). These issues related to the interpretation of quantum machanics are still often debated in the literature \cite{Ballentine, holland, aay}; they will not be discussed further in this paper.

However the derivation by Reginatto, though enlightening, also creates a puzzle: {\it Why is the inference method in statistical mechanics based on the information measure (\ref{kl}) while that used in obtaining Schrodinger's equation based on a different measure (\ref{fish})}? If, as discussed in \cite{reg2}, the use of the information measure (\ref{kl}) is well-motivated in statistical mechanics, then what is the motivation for using the Fisher information measure in the derivation of the Schrodinger equation? If there are several information measures, how does one decide {\it a priori} on which to choose? Is there a universal measure of which the various measures are special cases? These questions are investigated, to various degrees, in this paper.

It turns out that there is a close relation between the two measures (\ref{kl}) and (\ref{fish}) \cite{Kullback,reg2}. If in eq.(\ref{kl}) one chooses the reference  distribution $r(x)$ to be the same as $p(x)$ but with infinitesimally shifted arguments, that is $r(x) = p(x + \Delta  x)$, then to lowest order,
\begin{eqnarray}
I_{KL} ( p(x), p(x + \Delta(x)) &=& {- (\Delta x)^2  \over 2} I_F (p(x)) + O(\Delta x)^3  \label{connection}
\end{eqnarray}
So to lowest order, minimising the Fisher information is the same as maximising the relative entropy for two probability distributions that are  close to each other, $r(x) = p(x + \Delta x)$. In other words, to lowest order, one is indeed using the same information measure in both statistical mechanics and quantum theory, the main restriction in the quantum mechanics derivation being that one must compare a probabilty distribution with one that is slightly shifted. This restriction is not too difficult to motivate, for example one might guess that $\Delta x$ represents the intrinsic uncertainty in quantum mechanics measurements, or that it represents the resolution at which the coordinates become distinguishable (These possibilities will be revisited once estimates are made on $\Delta x$ later in the paper). 

But if one pursues this line of reasoning, then it suggests that the usual Schrodinger equation might only be an approximation, and that a generalisation would result if one used the full relative entropy given on the left hand side of ({\ref{connection}) rather than just its lowest order approximation that is the Fisher information measure: This is the essence of what is explored in this paper, leading to a nonlinear Schrodinger equation. As mentioned earlier, the quantum potential arises only from a variation of the information measure, so that changing the information measure will result in a modified quantum potential which is then the sole source of the deviation of the nonlinear Schrodinger equation from its linear counterpart. In particular the continuity equation (\ref{cont1}) is unchanged so that $p=\psi^{*} \psi$ is still conserved and has a sensible interpretation as probability density.

Although many nonlinear generalisations of Schrodinger's equation have been obtained before, a partial listing is in \cite{nonlin}, none has been motivated using information-theoretic arguments. On the other hand, various  information-theoretic arguments have in fact been used by some authors to {\it justify} the structure of conventional quantum mechanics within the context of certain axioms \cite{axioms}. Of course the axioms can be questioned and that is one avenue where nonlinearity might enter those approaches. Indeed in the paper by Summhammer there is just a hint of the posibility that the linear superposition principle might be an approximation.

The plan for the rest of the paper is as follows. In the next section the nonlinear Schrodinger equation is derived using the full Kullback-Liebler relative entropy for a single particle in one space dimension and its various properties discussed. Difficulties appear when the derivation is extended to higher space dimensions in Section (3) with the result that rotational invariance seems to be unavoidably broken. In Section (4) estimates are made on the length parameter which controls both the size of nonlinearities and also the scale of symmetry breaking. 
Also discussed in Section (4) are the links between causality and linearity and connections to other derivations of the Schrodinger equation in the literature.

In Sections (5,6) the Klien-Gordon and Dirac equations are examined within the information theoretic framework confirming the intimate link between nonlinearity and breaking of Lorentz symmetry. The relativistic equations are then used to study the possibility of effective energy dependent masses generated by nonlinearities and applications to neutrino physics discussed. 

The choice of information measure is re-examined in Section (7). Conditions that a physically acceptable information measure should satisfy for use in deriving quantum dynamics are discussed. The simplest measure satisfying all the conditions is a regularised version of the Kullback-Liebler measure containing one free parameter $\eta$ in addition to the length scale $L$ that controls the symmetry breaking and size of nonlinearities. The apparently privileged role played by the Fisher information measure in linear quantum theory is discussed in Section (8). 

The main results of this paper are summarised in Section (9) and some ways in which the nonlinear theory might be tested experimentally are highlighted. The paper ends with a discussion of various perspectives that one may take in deducing the laws of physics. A short appendix discusses quantum fields.

\section{Nonlinear Schrodinger equation for a single particle in one dimension}

The arguments of the last section invite one to use the distance measure  
\begin{equation}
\Phi_B \equiv -I_{KL}(p(x),p(x+L)) = \int p(x) \ln {p(x) \over p(x+L)} \ dx \ dt \label{kl2} \, ,
\end{equation}
instead of the Fisher information, and to minimise the action $\Phi_A + {\cal{E}} \Phi_B$ by varying both $S$ and $p$. Note that the measure (\ref{kl2}) is positive definite \cite{Kullback}, the Lagrange multiplier ${\cal{E}}$ has dimensions of energy and $L$ is a length parameter. The equations of motion that follow are 

\begin{eqnarray}
{\partial p \over \partial t} + {1 \over m} {\partial \over \partial x} \left(p {\partial S \over \partial x} \right)&=&0 
\, ,   \label{cont2} \\
{\partial S \over \partial t} + {1 \over 2m} \left( {\partial S \over \partial x} \right)^2 + V + Q_{1NL} &=& 0 \, ,  \label{hj2} 
\end{eqnarray}
where
\begin{equation}
Q_{1NL} = {\cal{E}} \left[ \ln {p(x) \over p(x+L)}  \ + 1 \ - {p(x-L) \over p(x)} \right] \, 
\end{equation}
is the generalised quantum potential that replaces $Q$. The two real equations can be combined into one complex nonlinear equation thorough the identification $\psi = \sqrt{p} \ e^{iS/ \hbar}$,
\begin{equation}
i \hbar {\partial \psi \over \partial t} = - {{\hbar}^2 \over 2m} {\partial^2 \psi \over \partial x^2} + V(x) \psi + 
F_1(p) \psi \, , \label{nsch1}
\end{equation}
with
\begin{eqnarray} 
F_1(p) &\equiv& Q_{1NL} - Q \\
&=& {\cal{E}} \left[ \ln {p(x) \over p(x+L)}  \ + 1 \ - {p(x-L) \over p(x)} \right] + {{\hbar}^2 \over 2m}  {1 \over \sqrt{p}} {\partial^2 \sqrt{p} \over \partial x^2} \label{fip}
\end{eqnarray}
and $p(x) = \psi^{\star}(x) \psi(x)$. The energy scale ${\cal{E}}$ and the length scale $L$ are not independent as we need to recover the usual Schrodinger equation to lowest order. That is, to lowest order ${\cal{E}} \Phi_B = {\hbar^2 \over 8m}I_F$ (or equivalently $Q_{1NL} = Q$ to lowest order) so that one obtains through eq.(\ref{connection}) the relation 
\begin{equation}
{\cal{E}} L^2 = {{\hbar}^2  \over 4 m } \,. \label{uncert}
\end{equation}
This equation has a natural interpretation. If a particle is localised to a region of size $L$ then its momentum would be at of order $\hbar /2L$ and its energy would be $\sim \hbar^2 /8mL^2$ which is consistent with (\ref{uncert}). Amusingly then, contained in the {\it kinematics} of the nonlinear Schrodinger equation is a a relation (\ref{uncert}) that is essentially the uncertainty principle of quantum mechanics. A critical point to note here is that if ${\cal{E}}$ is finite, then $L$  cannot be zero unless $\hbar$ vanishes. Thus the length scale is intimately connected to quantum mechanics. A more detailed discussion of the posible physical interpretations of ${\cal{E}}$ and $L$ follows in Section (4). 

The nonlinearity of the generalised Schrodinger equation  (\ref{nsch1}) is thus due to the nonvanishing of $F_1(p)$ which measures the deviation of the generalised potential $Q_{1NL}$ from the usual quantum potential $Q$. Expanding in $L$, one obtains
\begin{equation}
F_1(p) = {\hbar^2 L \over 4m} \left[ - { (p')^3 \over 3 p^3} + {p' p'' \over 2 p^2} \right] + O(L^2) \, , \label{f1est}
\end{equation} 
where the prime denotes derivative with respect to $x$. Interestingly, with appropriate boundary conditions, the space-time average of the new quantum potential $Q_{1NL}$, 
\begin{eqnarray}
\int p \ Q_{1NL} \ dx \ dt  = {\cal{E}} \int p(x) \ln {p(x) \over p(x+L)} \ dx \ dt 
\end{eqnarray}
is precisely the information measure used in the action $\Phi_A + {\cal{E}}\Phi_B$. This relation is the generalisation of that  found by Reginatto relating the average of the usual quantum potential to the Fisher information measure \cite{reg1}.

Here are some of the properties of the nonlinear Schrodinger equation (\ref{nsch1}):
\begin{enumerate}
\item The nonlinear equation still admits plane wave solutions, $\psi = e^{ikx-i\omega t}$ because $F_1(p=\mbox{constant})=0$. \\

\item If $\psi$ is a solution of the equation, then so is $\lambda \psi$ for any constant $\lambda$ because $F_1(\lambda^2 p) = F_1(p)$. This allows solutions to be normalised, the importance of which is discussed in Section (7).\\

\item The continuity equation (\ref{cont2}) still holds by construction so that $p(x)=\psi^{*} \psi$ still has a sensible interpretation as probability.\\

\item If two solutions $\psi^{(1)}$ and $\psi^{(2)}$ of the equation have negligible overlap then their superposition is also an approximate solution. This is because for negligible overlap, $F(p) = F(p^{(1)})$ {\it or} $F(p)=F(p^{(2)})$.  This ``weak superposition principle" ( Bialynicki-Birula et.al. \cite{nonlin}) will be used in a later section to constrain other possible information measures.

\item For stationary states that are normalisable, one gets by substituting $\psi \equiv e^{-iEt/\hbar} \phi(x)$ and averaging,
\begin{equation}
E = - {{\hbar}^2 \over 2m} <{\partial^2  \over \partial x^2}> + <V(x)>  + <F_1(p)>  \, , 
\end{equation}
so that the nonlinearity induced energy shifts are caused by an excess of energy in the modified quantum potential over the normal quantum potential. 

\item In the linear theory the kinematical content, represented by the state $\psi$, is separate from the  dynamics which is determined by the Hamiltonian. These two aspects become entwined in the nonlinear theory as the Hamiltonian becomes state-dependent through $F(p)$. 

\end{enumerate}

Most of these properties hold also for various other nonlinear equations that have been constructed in the past \cite{nonlin}. The potential singularities in (\ref{fip}) can be avoided by using a regularised measure to be discussed in Section (7.1).

\subsection{Parity preservation}

The nonlinear term $F_1 (p)$ is not invariant under the parity transformation $x \to -x$. However this is an artifact that is easy to remedy. Notice that in (\ref{connection})  $\Delta(x)$ can be either positive or negative but in (\ref{kl2}) and the sequel it was implicitly assumed that $L$ was positive. Therefore, in order to obtain a parity-invariant equation one should use a symmetrised form of Eq.(\ref{kl2}) involving both postive and negative values of $L$. Such a symmetrised measure is equivalent to the use in information theory of the $J$-divergence \cite{Kapur, Kullback}. 

The net result is the replacement 
\begin{equation}
F_1(p) \to {1 \over 2} ( F_{1}^{+}(p) + F_{1}^{-}(p) ) \label{symmp}
\end{equation}
where $F_{1}^{+}$ is the same as $F_1$ but $F_{1}^{-}$ uses $-L$ instead of $L$. Nevertheless, it is more useful for latter discussions to consider a {\it more general case whereby the $F_{1}^{\pm}$ are of the same form as $F_1$ but instead of $L$ they involve respectively  $\pm L^{\pm}$}, where $L^{\pm}$ are two different length scales. Then one has parity violation of order $(L^{+} - L^{-})$ with parity invariance only when $L^{+} = L^{-}$. The relation (\ref{uncert}) must be replaced in the general case by 
\begin{equation}
{\cal{E}} {(L^{+})^2 + (L^{-})^2 \over 2} = {{\hbar}^2  \over 4 m } \,. \label{uncertS}
\end{equation}

For brevity, in latter constructions and discussions the focus will be on simple measures like (\ref{kl2}) that violate parity maximally, as the more general case of parity conservation or small parity violation is easily obtained using a symmetrisation like (\ref{symmp}) with the symbols having a meaning as highlighted following that equation. In order to avoid possible confusion with other forms of parity violation, the parity violation due to the length parameters $L^{\pm}$ will sometimes be referred to as ``geometric parity violation". For the parity invariant case, one has instead of the estimate (\ref{f1est}) the next term of order $L^2$.     

There is no problem in extending the nonlinear Schrodinger equation  to more than one particle in one space dimension. However complications appear in moving to more than one space dimension even for a single particle. These are discussed in the Section (3).

\subsection{Periodic zero modes and localised states}

Consider the free ($V=0$) one-dimensional nonlinear Schrodinger equation (\ref{nsch1}) for a wavefunction $\psi = e^{-iEt/ \hbar} \phi(x)$ with $\phi$ real,   
\begin{eqnarray}
E &=& - {{\hbar}^2 \over 2m \phi } {\partial^2 \phi \over \partial x^2} + {\cal{E}} \left[ \ln {p(x) \over p(x+L)}  \ + 1  - {p(x-L) \over p(x)} \right] + {{\hbar}^2 \over 2m \sqrt{p} }  {\partial^2 \sqrt{p} \over \partial x^2} \label{zfirst}\\
&=&{\cal{E}} \left[ \ln {p(x) \over p(x+L)}  \ + 1 \ - {p(x-L) \over p(x)} \right] \, . \label{1dex}
\end{eqnarray}
The kinetic energy term (first term in (\ref{zfirst})) cancels with the quantum potential $Q$ (last term in (\ref{zfirst})) to leave only the modified quantum potential $Q_{1NL}$ as the sole determiner of the eigenvalue $E$. This result looks odd as  one is used to having a differential equation for the eigenvalue $E$, but note that if one expands $Q_{1NL}$ in the small parameter $L$ then the usual kinetic term reappears with higher derivative corrections!   

The eigenvalue equation (\ref{1dex}) has an exact solution: For {\it any} $\phi(x)$ that is periodic in $L$, $\phi(x+L) = \phi(x)$, the right-hand side vanishes and so $E=0$. Thus the nonlinear equation  supports stationary states that have zero energy and are periodic with a period equal to the length $L$ which sets the scale of nonlinearities. These periodic solutions exist for the exact equation (\ref{1dex}) but not for the perturbatively approximated equation obtained by expanding and truncating $Q_{1NL}$ to some power in $L$.

Nonlinear wave equations often have localised solutions. A detailed analysis of (\ref{nsch1}) needs to be performed to see if it supports solitary waves but here a heuristic discussion based on the quantum potential will be sketched. Consider a Gaussian wavepacket $\psi \sim \exp(- x^2/ \beta^2)$. The usual quantum potential  for this state is 
\begin{equation}   
Q = { \hbar^2 \over m \beta^2} \left( 1 - {2 x^2 \over \beta^2} \right)
\end{equation} 
which gives rise to a repulsive force, $-dQ/dx = \hbar^2 x / m \beta^4  >0$, that causes the wavepacket to disperse. The nonlinear quantum potential  $Q_{1NL}$ is equivalent to a force
\begin{equation}
- {d Q_{1NL} \over dx} = -  {2 {\cal{E}} L \over \beta^2} \left[1 - \exp\{-(L^2-2xL)/ \beta^2\} \right] \label{nonlinq}
\end{equation} 
which is attractive for small $x$. Thus there is a possibility that $F(p)$, which is an effective potential in the nonlinear Schrodinger equation, is attractive, potentially leading to nondisperive solutions. On dimensional grounds, it would appear that these solutions, if they did exist, would have an extension on the scale $L$.    

The discussion here has been for the case of maximal geometric parity violation. At the other extreme where parity is preserved, the zero mode solution still exists but the analog of (\ref{nonlinq}) gives a repulsive force. 

Assuming that a solution exists to the nonlinear equation whose form is slighly deformed from that of a plane wave, one can use dimensional analysis to obtain an estimate of the modification to the dispersion relation of a free particle. For the case of maximal geometric parity violation one obtains  from (\ref{f1est}),
\begin{equation}
F_1(p) \sim   {\hbar^2 L \over 4m } {1 \over L_{0}^3} \, ,
\end{equation} 
where $L_0$ is the characteristic scale in the problem such as the deBroglie wavelength $\hbar/k$. The symbol $k$ is used to denote momentum ({\it not} wavevector) to avoid confusion with $p$ which represents probability density. Hence the lowest order correction to the standard $k^2/2m$ relation might be expected to be $k^3 L / 4m \hbar$. On the other hand for the case where parity is preserved, one obtains $L^2 k^4 / 4 m \hbar^2$ as the correction to the standard dispersion relation. 
The length scale $L$ will be estimated in later sections.

\section{Several particles in many dimensions}

The Schrodinger equation for $N$ particles in $d+1$ dimensions is 
\begin{equation}
i\hbar \dot{\psi} = \left[ - {\hbar^2 \over 2} g_{ij} \partial_i \partial_j + V \right] \psi \, \label{schmulp}
\end{equation}
where $i,j= 1,2,......,dN$ and the metric is defined as $g_{ij} = \delta_{ij} /m_{(i)}$ with the symbol $(i)$ defined as   the smallest integer $\ge i/d$. That is, $i=1,...d$, refer to the coordinates of the first particle of mass $m_1$, $i=d+1,.....2d$, to those of the second particle of mass $m_2$ and so on. The overdot refers to a partial time derivative and the summation convention is used unless otherwise stated. As in Ref.\cite{reg1}, it is assumed that one is using Cartesian  coordinates as otherwise the interpretation of the term (\ref{fish}) in the variational action becomes unclear  within the information-theoretic context. On the other hand this is not an artificial restriction as the correspondence principle between observables such as momenta and their operator representation is unambiguous only in Cartesian coordinates \cite{Ballentine}.  

As usual, writing $\psi = \sqrt{p} \ e^{iS / \hbar}$ decomposes the Schrodinger equation into two real equations,
\begin{eqnarray}
\dot{p}  + g_{ij} \ \partial_i \left( p \partial_j S \right)  &=& 0 \, ,  \label{cont3} \\
\dot{S} + {g_{ij} \over 2} \partial_iS \partial_j S + V -{{\hbar}^2 \over 8} g_{ij} \left( {2 \partial_i \partial_j p \over p} - {\partial_i p \partial_j p \over p^2} \right) &=& 0 \, .  \label{hj3} 
\end{eqnarray}
The term with explicit $\hbar$ dependence in the last equation is the usual quantum potential. Equations (\ref{cont3},\ref{hj3}) follow from a variational principle \cite{reg1}: One minimises the action
\begin{eqnarray}
\Phi &=& \int  p \left[ \dot{S}  + {g_{ij} \over 2} \partial_i S \partial_j S   + V \right] dx^{Nd} \ dt   \ + \Phi_F \, \label{varmulp} 
\end{eqnarray}
where
\begin{eqnarray}
\Phi_F & \equiv & {{\hbar}^2 \over 8} g_{ij} \int dx^{Nd}  dt \  {\partial_i p \partial_j p \over p} \, .
\end{eqnarray}
is the relevant Fisher information measure for several variables. Since the metric $g_{ij}$ is diagonal, no mixed derivatives appear in the Fisher information $\Phi_F$. Recall that the diagonality of the metric is a consequence of the form of the kinetic term in the Schrodinger equation in Cartesian coordinates (\ref{schmulp}). Thus the structure of the measure $\Phi_F$ for several variables is fixed by the conventional kinetic term of the Schrodinger equation once one goes beyond one space dimension. The number of particles involved introduces no essential modification.

Following the strategy used in the previous section, one replaces the measure $\Phi_F$ by an appropriate generalisation of the Kullback-Liebler measure such that it reduces to $\Phi_F$ at lowest order. Consider first the attempt
\begin{eqnarray}
\int dx^{Nd} dt \ p(x_i) \ln {p(x_i) \over p(x_i + L_i)} & \approx& {L_i L_j \over 2} \int dx^{Nd} \ dt \  {\partial_i p \partial_j p \over p} \ + O(L^3) \, .
\end{eqnarray}
The lowest order term on the right-hand-side is not of the form $\Phi_F$. This is simply because the quadratic term in the Taylor expansion of a multivariate function generally contains mixed derivatives while, as emphasized above, $\Phi_F$ does not. Therefore it is necessary to arrange matters so that no mixed terms appear in the expansion of a suitably generalised Kullback-Liebler information measure. A little reflection shows that each coordinate degree of freedom must be treated separately. Consider therefore 
\begin{equation}
\Phi_{KL} \equiv \int dx^{Nd} dt \sum_{i=1}^{Nd} {\cal{E}}_i p(x) \ln {p(x) \over p_i(x)} \, \label{correct},
\end{equation}
where $p_i(x) \equiv p(x_1, x_2,.....,x_i + L_i,....,x_{Nd})$ means that the probability density has only its $i$-th coordinate shifted by a length parameter $L_i$ and where ${\cal{E}}_i$ are parameters with dimensions of energy. $\Phi_{KL}$ is therefore a  sum of measures, one for each independent degree of freedom. Expanding, one has 
\begin{equation}
\Phi_{KL} \approx \sum_{i} { {L_{i}}^2 {\cal{E}}_i \over 2} \int {(\partial_i p)^2 \over p} \ + \ O(L)^3 \, .
\end{equation}    
which is similar in form to $\Phi_F$. This similarity can be made into an identity by relating the length and energy parameters (as in the single particle case) 
\begin{equation}
{\cal{E}}_i L_{i}^{2} = {\hbar^2  \over 4 m_{(i)}}  \;\;\;\;  \mbox{(no sum on {\it i})} \, . \label{uncert2}
\end{equation}

Having identified the correct measure $\Phi_{KL}$, one can use it in (\ref{varmulp}) to replace $\Phi_F$. The variational principle then gives a quantum Hamilton-Jacobi equation which is equivalent to the following nonlinear Schrodinger equation 
\begin{eqnarray}
i\hbar \dot{\psi} &=& \left[ - {\hbar^2 \over 2} g_{ij} \partial_i \partial_j + V \right] \psi + F(p) \psi \, , \label{mpsch}  \\
F(p) & \equiv & Q_{NL} - Q \, , \\
Q_{NL} &=& \sum_{i}^{Nd} {\cal{E}}_i  \left[ \ln {p \over p_{+i}} + 1 - {p_{-i} \over p} \right] \, , \label{qnon} \\
Q &=& -{{\hbar}^2 \over 8} g_{ij} \left[ {2 \partial_i \partial_j p \over p} - {\partial_i p \partial_j p \over p^2} \right] \, , 
\end{eqnarray}
with
\begin{eqnarray}
p_{\pm i}(x) &=& p(x_1, x_2,.....,x_i \pm L_i,....,x_{Nd}) \,  
\end{eqnarray}
as defined after eq.(\ref{correct}). As before, the nonlinearity is determined by $F(p)$ which measures the deviation of the generalised potential $Q_{NL}$ from the usual quantum potential $Q$. Expanding in $L_i$, one obtains
\begin{equation}
F(p) = \sum_{i} {\hbar^2 L_i \over 4 m_{(i)}} \left[ - { (\partial_i p)^3 \over 3 p^3} + {\partial_i p \ \partial_{i}^2 p \over 2 p^2} \right] + O(L^2) \, . \label{noRot}
\end{equation} 

The $d+1$ dimensional multiparticle  nonlinear Schrodinger equation also has the properties listed for the single particle equation in the last section. Quite remarkably, the equation (\ref{mpsch}) is also separable: If $\psi = \psi^{(1)} \psi^{(2)}$ represents the composite wavefunction of two independent systems, $V= V_1 + V_2$, then it is easy to check that $F(\psi^{*} \psi) = F(\psi^{*(1)} \psi^{(1)}) + F(\psi^{*(2)} \psi^{(2)})$ so that the nonlinear Schrodinger equation decouples into two independent equations. 

It is important to note that separability was not demanded at the outset but is a {\it consequence} of requiring that  $\Phi_{KL}$ reproduce the Fisher measure $\Phi_F$ at lowest order for a {\it single} particle in higher dimensions. That simple requirement fixes the structure of $\Phi_{KL}$ even for the multiparticle case and the consequent separability of the equation. As in the one dimensional case, potential singularities in the equations of motion can be avoided by using a regularised measure to be discussed in Section (7.1).

Also, as for the one-dimensional case, the nonlinear term in equation (\ref{mpsch}) violates parity. The general situation is obtained by the replacement 
\begin{equation}
F(p) \to {1 \over 2} ( F^{+}(p) + F^{-}(p) ) \label{symmp2}
\end{equation}
in (\ref{mpsch}), where the $F^{\pm}$ are of the same form as $F$ but instead of $L_i$ they involve respectively  $\pm L_{i}^{\pm}$}. Then one has parity violation of order $(L_{i}^{+} - L_{i}^{-})$. Correspondingly the relation (\ref{uncert2}) gets replaced by          
\begin{equation}
{\cal{E}}_i { (L_{i}^{+})^{2} + (L_{i}^{-})^2 \over 2} = {\hbar^2  \over 4 m_{(i)}}  \;\;\;\;  \mbox{(no sum on {\it i})} \, . \label{uncertS2}
\end{equation}
In the parity symmetric case the lowest order term (\ref{noRot}) vanishes and the first surviving term is of order $L_{i}^2$.

\section{Lorentz symmetry breaking}

Even in the parity symmetric case, the nonlinear equation (\ref{mpsch}) is not rotationally, and hence Galilean, invariant. Therefore in the relativistic case the equation will not be Lorentz invariant as discussed in Section (5). This symmetry breaking is almost apparent from the form of $Q_{NL} $ where each coordinate is shifted separately, and is explicit in a perturbative expansion of $F(p)$. The rotational/Lorentz invariance is broken by the necessarily nonzero values of length parameters $L_i$ (even if they are the same), giving space(time) a cellular structure. 

There is currently no evidence for such a structure, or violation of spacetime symmetries, but several independent lines of enquiry \cite{kost,blum} have suggested that at sufficient high energies, probably close to the Planck scale, spacetime might not be smooth and our usual spacetime symmetries might be violated. 

Let us recapitulate the line of reasoning that has led us to the nonlinear Schrodinger equation with broken symmetries.  
It was driven by the desire to use a common  measure of information for inferring the proabilistic laws of physics. The relevant measure in the maximum entropy principle is the Kullback-Liebler information. Reginatto showed that the Fisher information measure was relevant for deriving the usual (linear) Schrodinger equation.  However the Fisher measure is an approximate case of the Kullback-Liebler information and using the latter measure gave rise to the nonlinear Schrodinger equation.  The breaking of continuous spacetime symemtries in the nonlinear equation then appears as an unavoidable consequence of using the Kullback-Liebler measure in the inference procedure.

In previous derivations \cite{nonlin} of nonlinear Schrodinger equations, Galilean invariance has been considered  a necessary constraint, just as symmetry constraints have traditionally been used in constructing Lagrangians in high energy physics. 
Indeed without a plausible alternative guiding principle, it would have been difficult to abandon symmetry constraints; see however \cite{origins}. In this paper  an inference principle has been used as the the main guide, leading to a nonlinear generalisation of the Schrodinger's equation and the consequent symmetry breaking. The results lead one 
to propose that the currently observed spacetime symmetries and the apparent linearity of Schrodinger's equation are two related phenomena and the potential future violation of one phenomenon  at small length scales should be accompanied by a violation of the other one too. Given the likelihood that our observable spacetime is an emergent structure, this then suggests that the linearity of conventional quantum mechanics will be broken at scales where there are deviations from the usual spacetime symmetries.
        
In passing, it is noted that in low energy physics applications where a nonlinear quantum equation might be used for phenomenological reasons, the symmetry breaking might be natural in some circumstances.

\subsection{Scale of Lorentz symmetry breaking} 

As noted in the discussion following (\ref{uncert}), the length $L$ is dependent on $\hbar$ and so it
is tempting to identify $L$ with the Planck Length\footnote{The explicit powers of $\hbar$ do not match  in $L$ and $L_p$ but there could be some implicit dependence in $L$.}, $L_p =(\hbar G / c^3 )^{1 \over 2} \sim 10^{-35}$m. However let us consider first various alternatives with larger length scales.

As discussed earlier, the relations (\ref{uncert}, \ref{uncert2}) have the form of the uncertainty principle for a particle
localised to a region of size $L$. Therefore one possibility to explore is whether  ${\cal{E}}$ might be some energy intrinsic to the particle. If one chooses 
${\cal{E}} \propto m c^2$ then from  (\ref{uncert}) $L$ is essentially the Compton wavelength $\lambda_c =\hbar/mc$. However this conflicts with the hydrogen atom spectrum: Consider the energy correction to  ground state of the hydrogen atom. Since the ground state is spherically symmetric, the perturbative energy correction $<F(p)>$ calculated using (\ref{noRot}) vanishes because of the odd number of derivates in the $L^3$ term. Hence the lowest perturbative correction would be the next term of order $L^4$ giving, on dimensional grounds,
\begin{equation}
\Delta E \sim {\cal{E}} \left( { L \over a_o} \right)^{4} \label{pert4}
\end{equation}  
where $a_0 = \hbar^2 /m e^2$ is the Bohr radius $\sim 10^{-10}$m. Since the ratio $\lambda_c / a_0 = \alpha = 1/137$ is the fine-structure constant, therefore $\Delta E \sim mc^2 \alpha^4$ which is of the same order as the fine structure correction to the hydrogen atom, and so much too large. Thus ${\cal{E}}$ cannot be the rest mass energy of  a particle.

Another possibility is that $L$ is the size of fundamental particles to which strict quantum mechanical considerations apply. The classical electromagnetic radius of an electron is $R_e = e^2/mc^2 \sim 10^{-15}$m. Using $L = R_e$ and combining (\ref{uncert2}) with (\ref{pert4}) gives $\Delta E \sim mc^2 \alpha^6$ which is somewhat smaller than the Lamb shift and hyperfine structure corrections. The actual size of the electron, if it has any well-defined extension at all, is smaller than $10^{-18}$m and so there is a remote possibility that $L$ might be related to the size of elementary particles. 

Various low-energy atomic physics experiments searching for signatures of Lorentz violation have determined that the energy parameters in effective lagrangians measuring deviation from Lorentz invariance \cite{kost} must be tiny \cite{blum}, ``typically" of order  $10^{-27}$ GeV. One can now make a crude estimate for $L$ by assuming 
$\Delta E$ in (\ref{pert4}) to be $10^{-27}$ GeV and using the  relation (\ref{uncert}) for the electron. This gives  $L \sim 10^{-20}$m which is much larger than the Planck length of $10^{-35}$m. Of course as experiments get better the estimate might get pushed to the Planck value after all. Other estimates for $L$ are discussed in a later section.

Bear in mind that the nonlinear Scrodinger equation of this paper  suggests a {\it simultaneous} violation of both the linearity of quantum mechanics and a violation of spacetime symmetries. These entangled effects might complicate, in the present context, the interpretation of experiments that have been searching for only one violation or the other. Some experiments that are searching only for Lorentz violation have been mentioned above. There have also been  experiments searching for nonlinearities in quantum mechanics (assuming the usual spacetime symmetries). However most of these experiments have been motivated by Weinberg's particular class of equations \cite{nonlin} to look at spin systems. The nonlinearity  that has been found in this paper is related however to spatial variations in the wavefunction and so more experiments of such type would be welcome. Nevertheless if spacetime symmetries are indeed violated, then this should be taken into account  in the effective Hamiltonains used for interpreting the data of the experiments in \cite{atom}. That is, it might be useful to reanalyse the past data in the context of possible Lorentz violation. Interestingly, the limits on energy parameters \cite{atom} describing possible nonlinear deviations in quantum mechanics (assuming spacetime symmetries hold) are roughly the same order of magnitude as the experiments searching only for Lorentz violation in \cite{kost,blum}. This coincidence probably  simply reflects the current sensitivities of the experiments. 
 
When should one expect the nonlinear quantum corections to become large?  Roughly, this should happen when the length scale $L$ of the nonlinearity is comparable to the relevant length scale of the linear theory. Taking for simplicity $R_e$ for $L$, then for the hydrogen atom the nonlinear effects should be large when $R_e \sim a_0$ which reduces to $\alpha^2 \sim 1$, that is, in the regime where the running fine structure constant is large. More generally one would expect the nonlinear corrections to be large in electrodynamics when $L \sim R_e$ is comparable in size to the Compton wavelength of the electron $\lambda_e \sim \hbar/mc$, leading to $\alpha \sim 1$, again in the strong coupling regime. 

In short, $L$ might conceivably be much larger than the Planck length and perhaps related to the size (if any) of elementary particles. Indeed, what the form of the Kullback-Liebler relative entropy (\ref{connection}) is possibly suggesting is that $L$ is the resolution at which the spatial coordinates become distinguishable --- and this might be due either to the fuzziness of space at short distances or a small but finite extension of elementary particles, or both.

The discussion so far has assumed that the length parameters $L_{i}$ are constant, and for simplicity one may even take them to be the same $L_{i}=L$. However a more general viewpoint would be to consider the parameters to be dependent on position, $ L_i(x)$, and not just on the particular axis in configuration space that is chosen. The discussion can proceed as before if the $L_i (x)$ are kept inside the integrals and, for example for a single particle of mass $m$, the constraint  (\ref{uncert2}) generalised to 
\begin{equation}
{\cal{E}}_i L_{i}^{2}(x) = {\hbar^2  \over 4 m}  \;\;\;\;  \mbox{(no sum on {\it i})} \, . \label{uncert22} \,
\end{equation}
This requires for consistency that $\hbar^2 /m$ be space dependent\footnote{Another possibility is to promote the Lagrange multiplier ${\cal{E}}$ to a background field ${\cal{E}}(x,t)$.}. However for the Schrodinger equation  to be derivable from a variational action of the form (\ref{varmulp}), $\hbar$ should be kept constant. With a position dependent mass 
$m(x)$ and a symmetrised form for kinetic energy operator, $\hat{p} {1\over 2 m(x)} \hat{p}$, the derivation of Ref.\cite{reg1} goes through as discussed in \cite{plast}. Then the nonlinear generalisation as discussed in this paper can take place. 

Position dependent masses might be relevant for describing the effective dynamics at short distance scales where space is not expected to be smooth. In Sections (5,6) it will be seen that in the relativistic case quantum nonlinearities can sometimes mimic effective spacetime dependent masses even when the $L_i$ are constant and there are no explicit mass terms. Also, in the relativistic case, one can make $L_i$ time dependent if one so desires.

\subsection{Causality and Linearity}

Weinberg's general study of nonlinear quantum mechanics stimulated a large amount of research into the subject. Some of the early theoretical investigations concluded that nonlinear quantum mechanics would allow unphysical effects such as superluminal signals \cite{gisin}. However 
those studies implicitly assumed for nonlinear quantum mechanics some of the usual structure of the linear theory. Within a larger framework a nonlinear theory might be consistent after all \cite{mielnik}. On the other hand, working within certain assumptions, other  studies \cite{Svet} have suggested that there is indeed some connection between causality and the linearity of quantum mechanics.

The nonlinear theory obtained in this paper explicitly violates spacetime symmetries, and it has been argued that this is indeed unavoidable within the information theoretic framework. Thus the conclusions of this paper are qualitatively consistent with previous studies. In particular, since the spacetime symmetries have been explicitly broken, then any possible effects due to such violations cannot be considered pathological unless ruled out by experiments.

\subsection{Connections}

Over the years there have been many approaches to understanding the usual linear Schrodinger equation. In Nelson's \cite{Nelson} derivation, quantum mechanics is the result of a type of Brownian motion while in Smolin's work \cite{Smolin}, Nelson's basic idea is married to matrix models. As one can see from Ref.\cite{Smolin}, the extra term that is obtained there because of the nondifferentiability of paths is precisely the usual quantum potential, which as noted in Ref.\cite{reg1}, and as discussed above, is equivalent to the Fisher information measure in the variational action. Thus a generalisation of these works to yield nonlinear quantum mechanics of the type considered in this paper would mean that deviations from Brownian  motion must be considered. Of course in the model of \cite{Smolin, Smolin2} nonlinearities of a different kind might also arise when some of the approximations in those derivations are relaxed.

Another recent derivation of the linear Schrodinger equation used the idea of adding fluctuations to the classical Hamilton-Jacobi equation \cite{hall} and then constraining those fluctuations through an exact uncertainty relation \cite{hall2}. Mathematically, the result is again the emergence of the usual quantum potential and thus the linear Schrodinger equation. In this approach a generalisation to a nonlinear equation might be possible through the use of generalised uncertainty relations.  

Studies of nonlinear quantum mechanics in the literature typically assume that the usual spacetime symmetries are still valid while studies of spacetime structure at small scales typically assume that the usual quantum mechanics is valid. There have been a few speculations that both linear quantum mechanics and the large scale structure of spacetime might be emergent quantities derivable from a more fundamental theory, see for example Ref.\cite{Penrose}. 
In \cite{Smolin}, it was also suggested that both macroscopic spacetime and linear quantum mechanics might have a common origin. The equation (\ref{mpsch}), derived in this paper using an information theoretic principle, displays explicitly a link between spacetime symmetries and the linearity of quantum mechanics and so might be an intermediate candidate for the abovementioned search for  a unified theory of quantum matter and spacetime.

A word on spin. One may also decompose the Pauli equation in terms of hydrodynamical variables \cite{holland}. 
There appears now in addition to the quantum potential $Q$ of the spinless Scrodinger equation, another quantum potential $Q_s$ related to spin degrees of freedom. The nonlinearisation of $Q$ can proceed as before, but $Q_s$ has no obvious connection with information theoretic measures.  The resulting nonlinear Pauli equation would therefore contain the same extra term $F(p)$ as the spinless Schrodinger equation. A different approach would use the result of Ref.\cite{reg3} where Reginatto combines $Q$ with $Q_s$ and relates the result to a generalised Fisher information. One might attempt to nonlinearise that equation using an information theoretic approach. A third approach is discussed in Section (6).

\section{Klien-Gordon equation}
The single particle Klien-Gordon equation was also derived in Ref.\cite{reg1}. The nonlinear extension is discussed here with some change in notation from Ref.\cite{reg1}. 

The Lagrangian density for the classical single-particle Klein-Gordon field $\phi$ is written as 
\begin{equation}
{\cal{L}} =  \hbar^2 \partial_\mu \phi \partial_\mu \phi^{\star} + {m^2c^2 } \phi \phi^{\star} \, \label{kglag}
\end{equation}
which upon variation of the action gives the equation of motion
\begin{equation}
 -\hbar^2 \partial_\mu \partial^{\mu} \phi + {m^2c^2} \phi  =0 \, . \label{kgeom}
\end{equation}
the metric is $(-,+,+,+)$, $x_0 =ct$. The normalisations have been chosen so that when $\phi = \exp[-imc^2t/\hbar] \ \psi /2m$ the nonrelativistic limit of (\ref{kgeom}) is the usual Schrodinger equation. 
Now substitute in $\phi = \sqrt{p} e^{i S/\hbar}$ into (\ref{kglag}) to get 
\begin{equation}
{\cal{L}} = p\left( { (\partial_{\mu} S)(\partial^{\mu} S)} + {m^2c^2}\right) + {\hbar^2 \over 4 p} (\partial_{\mu} p)(\partial^{\mu} p) \,. \label{kglag2}
\end{equation}
Varying the action $\int dx^3 dx_0 {\cal{L}}$ with respect to $S$ gives the continuity equation $\partial_{\mu}(p \partial_{\mu} S)=0$ which implies the well known fact that $p=\phi \phi^{\star}$ is not the conserved quantity even though it is positive definite. Although $p(x)$ is not a probability density, it will be used as a measure of likelihood so as to give the last term of (\ref{kglag2}) the interpretation of a Fisher information density. 

In the information-theoretic approach of \cite{reg1} there is a neat separation in the equations of the classical part from the quantum fluctuations that are later inserted through the principle of maximum entropy. Such a procedure is quite clear for nonrelativistic dynamics but some ambiguities in interpretation and other technical difficulties appear in trying to apply the procedure to relativistic systems and spin degrees of freedom. The following  additional rule is adopted to guide one in the various constructions: A generalised nonlinear equation at a higher level (e.g. relativistic) should reduce to the nonlinear equation at the lower level (non-relativistic) in some limit (e.g. large $c$.) The use of   
$p=\phi \phi^{\star}$ is an example of this as we shall see. 

Thus one replaces the Fisher measure (the space-time integral of the last term of (\ref{kglag2}))  by the  more general KullBack-Liebler information measure which is similar in form to the one used in the non-relativistic case,
\begin{equation}
 \int dx^{3} dx_0 \sum_{l=0}^{3} {\epsilon_l} \ \beta_l \ p(x) \ln {p(x) \over p_{l}(x)} \, \label{relat},
\end{equation}
where as before $p_{l}(x)$ means that $p(x)$ has only its $l$-th spacetime coordinate shifted by a length parameter $L_{l}$ and where ${\epsilon}_l$ are parameters with dimensions of energy times mass. The parameters $\beta_l = (-1,1,1,1)$ are the diagonal elements of the metric. The form (\ref{relat}) ensures that to lowest order in $L_l$ the usual Klein-Gordon equation will be recovered.  Furthermore noting that $x_0=ct$ and that $x_0 + L_o \to t + L_o/c$ then in the nonrelativistic limit (\ref{relat}) reduces to the Kullback-Liebler measure (\ref{correct}) and the Schrodinger equation is recovered as  required. The  parameters $\epsilon_l$ and $L_l$ are related by 
\begin{equation}
{\epsilon}_l L^{2}_{l}= {{\hbar}^2  \over 2 } \;\;\;\;  \mbox{(no sum on {\it l})} \, . \label{quncert2}
\end{equation}
Compared to the nonrelativistic case (\ref{uncert}) the mass does not appear in this relation because of the use of the conventional form of the Klien-Gordon lagrangian (\ref{kglag}); its dimensions  have been absorbed into the definition of $\epsilon_l$. 

As before, the equations of motion that follow from the generalised Lagrangian are nonlinear with the nonlinearity due to an excess of the new quantum potential over the usual quantum potential: 
\begin{equation}
-\hbar^2 \partial_\mu \partial^{\mu} \phi + {m^2c^2} \phi  + {\cal{F}}(p) \phi = 0 \, , \label{kgn}\\
\end{equation}
with
\begin{eqnarray}
{\cal{F}}(p) & \equiv & {\cal{Q}}_{NL} - {\cal{Q}} \, , \\
{\cal{Q}}_{NL} &=& \sum_{l=0}^{3} {\epsilon}_l \beta_l \left[ \ln {p \over p_{+l}} + 1 - {p_{-l} \over p} \right] \, , \\
{\cal{Q}} &=& -{{\hbar}^2 \over 4} \sum_{l=0}^{3} \beta_l  \left[ {2 \partial_l \partial_l p \over p} - {\partial_l p \partial_l p \over p^2} \right] \, , 
\end{eqnarray}
Expanding in $L_l$, one obtains
\begin{equation}
{\cal{F}}(p) \approx \sum_{l=0}^{3} {\hbar^2 \over 2} \beta_l {L_{l}} \left[ - { (\partial_l p)^3 \over 3 p^3} + {\partial_l p \ \partial_{l}^2 p \over 2 p^2} \right] + O(L^2) \, . \label{noRot2}
\end{equation} 

The equations above violate parity and time-reversal invariance maximally because of the nonlinear correction which is asymmetric 
in $L_l$. The general situation is obtained after the replacement 
\begin{equation}
{\cal{F}}(p) \to {1 \over 2} ( {\cal{F}}^{+}(p) + {\cal{F}}^{-}(p) ) \, , \label{symmp3}
\end{equation}
where the ${\cal{F}}^{\pm}$ are of the same form as ${\cal{F}}$ but instead of $L_l$ they involve respectively 
 $\pm L_{l}^{\pm}$. Then one has parity and time-reversal violation of order $(L_{l}^{+} - L_{l}^{-})$. Correspondingly the relation (\ref{quncert2}) gets replaced by   
\begin{equation}
{\epsilon}_l ( (L^{+}_{l})^2 + (L^{-}_{l})^2 ) = {{\hbar}^2 } \;\;\;\;  \mbox{(no sum on {\it l})} \, . \label{quncertS}
\end{equation}
       
In the parity and time symmetric case the lowest order term (\ref{noRot2}) vanishes and the first surviving term is of order $L_{l}^2$.

The main result of using an inference method with the help of the Kullback-Liebler measure is now a nonlinear Klien-Gordon equation without Lorentz invariance\footnote{However space and time translation symmetries are still preserved.}, just as rotational invariance was broken in the non-relativistic equation. This is easy to see in a perturbative expansion in $L_i$ of the Kullback-Liebler measure. Thus space and time attain a cellular structure. However just as in the nonrelativistic case, plane waves with the usual dispersion relation exist as solutions to the nonlinear equation in addition to other possible states with modified dispersion relations. 

One may interpret the nonlinear term ${\cal{F}}(p)$ in the equation of motion as an effective spacetime dependent mass generated by self-interactions. However since ${\cal{F}}(p)$ vanishes when $p$ is a constant, this possibility  does not exist for plane waves $\phi = e^{ik.x}$. Assuming a deformation of a plane-wave is a solution, then one can estimate the contribution of nonlinearities to an effective mass using dimensional analysis. Let us call this $\Delta m_0$, then for the case of maximal geometric parity violation, from (\ref{kgn}, \ref{noRot2}),
\begin{eqnarray}
\Delta m_{0} &=&  \left( { {\cal{F}}(p) \over  c^2 } \right)^{1 \over 2} \\
& \sim & {\hbar  \over c L} \left( {L \over L_0} \right)^{3 \over 2} \, \label{mass1}
\end{eqnarray} 
where $L_0$ is some characteristic length scale in the problem such as the Compton wavelength, $\hbar/m_0 c$. In terms of masses the relation becomes,
\begin{equation}  
{ \Delta m_0 \over m_0} \sim \left({m_o \over M}\right)^{1 \over 2} \, \label{massEst}
\end{equation}
where $M$ is the mass scale corresponding to the Lorentz symmetry breaking length scale $L=\hbar/Mc$. The current 
\cite{nist} relative uncertainty in the electron mass is $\sim 10^{-7}$. Using this for the left-hand-side of (\ref{massEst}) implies the lower bound estimate $M \sim 10^{11}$ GeV corresponding to $L \sim 10^{-27}$m . The proton mass has a similar relative uncertainty and if one assumes that the equation may also be used for composite objects then a lower bound estimate of $M \sim 10^{14}$ GeV is obtained. 
This crude analysis suggests that the scale of a possible simultaneous parity and Lorentz symmetry breaking might be close to the GUT scale. In this respect, precision measurements of the masses of elementary particles, and their possible momentum dependence might be useful in determining the scale $L$. 

On the other hand for the case where there is no geometric parity violation, but only Lorentz violation, one has 
\begin{equation}  
{ \Delta m_0 \over m_0} \sim \left({m_o \over M}\right) \, \label{massEst2} ,
\end{equation}
which results in the lower bound mass scale estimates mentioned above being reduced by a factor of $10^{-7}$.

Instead of using the Compton wavelength for $L_0$, one might use the deBroglie wavelength, $\hbar/k$, and so obtain a modification to the dispersion relation of a relativistic particle due to nonlinearities. When there is maximal geometric parity violation one obtains from (\ref{noRot2}) ,
\begin{eqnarray}
{\cal{F}}(p) &\sim& {\hbar^2 L \over L_{0}^3} \, \\
&\sim& { k^3 \over Mc} \, ,
\end{eqnarray}
as the correction to be added to the right-hand side of the usual $E^2 / c^2  = k^2 + (mc)^2$. In the parity symmetric case the relevant correction is $k^4/M^2 c^2$.

The massless limit of the nonlinear equation (\ref{kgn}) is even more interesting as now the nonlinear corrections would imply some effective mass-like behaviour; this is also true for the ultrarelativistic limit of low mass particles. The analysis is presented below for neutrinos.

\section{Dirac and Pauli equations}
The Fisher information measure, and the other information measures that contain it, are clearly related to quadratic kinetic terms as seen in Sections (1,2). On the other hand the Dirac equation is of first order. Therefore a provisional approach  will be adopted to establish a nonlinear generalisation of the Dirac equation\footnote{A different approach will be discussed elsewhere \cite{parwani}.}: Following Dirac, the relevant Hamiltonian is defined to be the square-root of the Klien-Gordon Hamiltonian even in the nonlinear case. Thus
\begin{equation}
i \hbar \dot{\psi} \equiv \sqrt{\hat{k}^2 + m^2 + \breve{F}(p)} \ \psi   \, \label{definedirac}
\end{equation} 
where $\breve{F}(p)$ is an appropriate matrix generalisation of the nonlinearity ${\cal{F}}(p)$. For example, one may take 
$\breve{F}(p)$ to be a diagonal matrix with elements ${\cal{F}}(p_a)$ corresponding to the $a$-th spinor component. In this way each spinor component will independently satisfy the nonlinear Klien-Gordon equation at least when $\breve{F}(p)$ is time-independent. Note that $p$ is the probability density, not to be confused with the momentum which has been  denoted by $k$. Since for all practical applications the nonlinear terms will be small, then to a very good approximation, 
\begin{equation}
i \hbar \dot{\psi} \approx \left[ (\alpha \cdot \hat{k} + \beta m) + {\breve{F}(p) \over 2 (\alpha \cdot \hat{k} + \beta m)} \right]  \psi   \, , \label{approxdirac}
\end{equation}
with $\alpha, \beta$ the usual Dirac matrices.

The advantage of defining the nonlinear Dirac equation in this manner is that the link with  nonlinear Klien-Gordon equation and also the usual Dirac equation is manifest. Furthermore, one may define the nonlinear Pauli equation as the nonrelativistic limit of (\ref{approxdirac}). Such a nonrelativistic Pauli equation might be relevant for interpreting results of low energy atomic physics experiments searching for nonlinearities or Lorentz violation \cite{blum,atom}.     

In the nonlinear Dirac equation one may approximately identify the positive-definite quantity  $\psi^{\dagger} \psi$ with the probability density as from (\ref{approxdirac}) one deduces that the usual conservation law is violated only by the small nonlinearities. Thus as usual there is an improvement over the Klien-Gordon equation. However since the probability density is not a Lorentz scalar in the Dirac case, its use in the nonlinear term $\breve{F}(p)$ contributes additionally to the Lorentz violation that is already there.   

Potential applications are likely to involve coupling to background gauge fields. This can be done as usual through the prescription $\hat{k} \to \hat{k}_{\mu} - e A_{\mu}$; the nonlinear term $\breve{F}(p)$ is gauge-invariant. The gauge-invariance of the equation can then be used to establish conservation laws if required.

\subsection{Neutrino oscillations}
In the minimal Standard Model all particles except the neutrino obtain a mass through condensates. However the observation of neutrino oscillations, see for example \cite{neutrinoM} for a review, has suggested that neutrinos might have a mass much smaller than the other particles. It is usually stated that the origin of this mass is due to physics beyond the Weinberg-Salam model. An alternative scenario for the phenomena is suggested here, involving physics beyond linear quantum theory.

The neutrino's equation of motion is postulated to be of the form  (\ref{approxdirac}). Then even if there is no explicit mass term, $m=0$, the nonlinearities will in general affect the propagation of neutrinos, thus possibly mimicing a energy-dependent mass. Thus it is suggested that the neutrinos are strictly massless, at least at the scale of the current experiments, and the apparent mass is due to quantum nonlinearities. These nonlinearities are naturally tiny since they are tied to Lorentz violation. For a crude estimate, use (\ref{mass1}) and identify $L_0$ with the deBroglie wavelength $\hbar/k$ of a massless neutrino propagating as a slightly deformed plane wave. Choose, for example, $L$ to be the GUT scale. Let $ E = k c$, then the effective energy-dependent mass of the neutrino in the case of maximal geometric parity violation is
\begin{equation}
\Delta m_o c^2 \sim 10^{-2} \ ( E \ \mbox{MeV})^{3 \over 2} \;\; \mbox{eV} \,. \label{Nmass} 
\end{equation} 
In (\ref{Nmass}) the energy of the neutrino is measured in MeV and the mass value is quoted in eV. For example, for $E=100$ MeV, the result is $10$ eV. If the Planck scale is used instead of the GUT scale then another suppression of $10^{-2.5}$ would result. Indeed the induced mass varies with the symmetry breaking mass scale as $1/\sqrt{M}$, which is unusual compared to conventional ``physics beyond the standard model" scenarios which assume quantum linearity. However for the case where there is no geometric parity violation, the effective mass scales as $1/M$ and at the GUT scale it is 
\begin{equation}
\Delta m_o c^2 \sim 10^{-11} \ ( E \ \mbox{MeV})^{2} \;\; \mbox{eV} \, , \label{Nmass2} 
\end{equation} 
which is much smaller than (\ref{Nmass}).

Current direct measurements on neutrino masses have only placed upper bounds in the eV range, thus the actual mass could indeed be zero, with ambiguities caused by the unusual momentum dependence discussed here. It would be interesting to do a careful analysis of the energy dependence of the observed neutrino `masses" and compare with the $E^{3/2}$ behaviour suggested above for the parity violating case. Note that for consistency, the perturbative estimate (\ref{mass1}) is valid only when $L \ll L_o$. For a $100$ MeV neutrino, $L_0 \sim 10^{-15}$m, so the condition is abundantly satisfied for $L$ at the GUT or Planck scale.

A mass-like term in the equation of motion of the neutrino could then explain, using conventional methods, the oscillations that are observed. However a more natural and simpler explanation is that the oscillations are due to the nonlinear terms coupling different neutrino species. In the minimal Standard Model the neutrinos of different flavours are arranged simply in a multiplet, $\psi = (\psi_e, \psi_{\mu}, \psi_{\tau})^{T}$, for example in the neutral current interaction. So it is natural to choose $\breve{F}(p)$ in (\ref{approxdirac}) such that $p = \psi^{\dag} \psi$ rather than the form mentioned earlier after (\ref{definedirac}).  This then couples the evolution of all the neutrino flavours in (\ref{approxdirac}) so that even if initially only one flavour is present, it will seed the growth of another while it itself decreases, thus resulting in oscillations. 

The details of the above proposal have to be worked out \cite{parwani} but one sees an obvious consequence of the above coupling: There is now a discrete (permutation) symmetry relating the neutrino flavours as they appear symmetrically in $p= \psi^{\dag} \psi$ and hence $\breve{F}(p)$. At scales where the nonlinearities are negligible the neutrinos would appear  decoupled and the symmetry would be vacuous. This permutation symmetry thus seems to shed some light on the mystery of family replication. 

Remarkably, a search of the literature reveals that some of the qualitative but almost inevitable possibilities that have been mentioned above have also been arrived at by other authors,  using very different motivations and arguments. The possibility of neutrino oscillations being due to Lorentz violation has been studied by several authors independently, see for example the references in \cite{kost2}. Kostelecky and Mewes have classified the possible Lorentz violating terms that can appear in an effective Lagrangian. Unlike this paper, they and others of course assume that linear quantum theory is valid even though Lorentz symmetry has been violated. Nevertheless a qualitative comparison reveals some striking similarities:  In \cite{kost2} it was noted that novel and unconventional features are possible for neutrino oscilaltions caused by Lorentz violation, such as direction dependence and unusual energy dependence. These features are explicit in the equation of motion (\ref{approxdirac}).

The approach of \cite{kost2} was to consider the most general effective Lagrangian that could describe neutrino physics in the presence of Lorentz violation. As a result they have many free parameters and possibilities. Here a definite proposal for the origin of Lorentz violations has been made, tied to quantum nonlinearities, and so hopefully the possibilities can be narrowed down for comparison with experiments.   

A possible discrete family symmetry in the Standard Model has also been noted by other authors through a study of the empirical neutrino mixing matrix \cite{mixing}. Ironically we have been led to a possible discrete family symmetry through a study of the nonlinearity in the Lorentz {\it symmetry violating} Dirac equation.

\section{Uniqueness}
The Kullback-Liebler information measure satifies certain axioms that are considered suitable for an information measure \cite{Shannon, max, Kapur}.  Of course if one relaxed some of the assumptions then other information measures might be admittable. Here let us consider the possibility of constructing nonlinear quantum mechanics from some other common information measures which reduce, in the limit when some parameter becomes small, to the Fisher information measure,  so that we are sure to recover the usual linear Schrodinger equation in that limit.

One interesting measure is the Renyi relative entropy \cite{Kapur},
\begin{equation}
R^{\alpha}(p(x),r(x)) = {1 \over 1 - \alpha} \ln [\int dx p^{\alpha} r^{1-\alpha}] \, .
\end{equation}
As $\alpha \to 1$ this reduces to the Kullback-Liebler measure while using $r(x) = p(x+L)$ and taking $L \to 0$ gives $-\alpha L^2 I_F (p) /2$. Another measure is the Wooters measure \cite{Wooters}
\begin{equation}
W(p(x),r(x)) = \cos^{-1} [ \int dx \sqrt{pr} ] \, .
\end{equation}
For $r(x)=p(x+L)$ one obtains $W^2 \to L^2 I_F(p)/4$ as $L \to 0$. Thus one may potentially use either $R^{\alpha}$ or $W^2$ in the variational action and obtain a nonlinear Schrodinger equation. Separability of the equation for multiparticle states will be automatic when one uses a sum of information measures, one for each coordinate degree of freedom as discussed in the text. However both the Renyi measure and the (square of) the Wooters measure create problems. Firstly, the measures are not integrals over a density as one would like for an action. So the equations of motion that follow will be highly nonlocal, containing integrals, and the weak superposition principle will not hold: States with negligible overlap will influence each other strongly. 

Another drawback with the resultant nonlocal equations is that they will not be homogeneous, that is, scaling a solution of the equations by a constant does not give another solution so that normalisation and interpretation of states after a measurement process would be problematic, as discussed by Kibble \cite{nonlin}. For the equations to be homogeneous, the measures used for the variational action (in the $P,S$ variables) should scale like $\lambda^2$ when $\psi \to \lambda \psi$. Note that here we have examples of equations that are not homogeneous yet are separable, in contradistinction to some other scenarios \cite{nonlin}.

If one is looking for deformations of the linear Schrodinger equation, then the weak superposition principle and homogeniety are desirable criteria as they allow for easy interpretation of solutions. Therefore according to those criteria the information measures above can be eliminated. On other hand if one were interested in more general dynamics that might potentially be relevant at short distance scales then these and many other information measures might still be valid.

Thus here we are interested in an information measure which satisfies the following conditions:
\begin{itemize}

\item $[C1] \;$ The information measure should be positive definite and of the form $G(p(x); L) = \int p \ H(p ; L) \ dx dt $ where $H$ is a function of the probability $p(x,t)$ and is $L$ some parameter.   This form ensures the validity of the weak superposition principle in the equations of motion. 

\item $[C2] \;$ $H$ should be invariant under scaling, $H(\lambda p ; L) = H(p ; L)$. This allows solutions to be (re)normalised.

\item $[C3] \;$  $G(p ; L\to 0)$ should give the Fisher information measure at lowest order. This allows  the linear Schrodinger equation to be recovered as an approximation. As discussed in the text, separability of the multiparticle equation is a secondary consequence of this condition when an appropriate sum of measures is used to reproduce the correct form of the linear Schrodinger equation in the limit $L\to 0$.
\end{itemize}

Are there any information measures that satisfy the conditions [C1-C3] other than the already discussed Kullback-Leibler distance? (It must be emphasized that  one is interested in genuine {\it information} measures which have some reasonable interpretation and properties \cite{Kapur} rather than arbitrary functions which only satisfy the above listed conditions, of which there are an infinite number). 

Indeed there are a quite a few, the best known probably being a ``q-deformation" of the Kullback-Liebler distance. A generalised entropy that has been much studied in recent years is Tsallis entropy, independently discovered earlier by Havrda and Chavrat \cite{HC},
\begin{equation}
I_{GS}^{(q)} = {1 \over 1-q} \int p^q (1-p^{1-q})  \, .
\end{equation}

It has been used to describe nonextensive statistics and is formally a modification of the usual Gibbs-Shannon measure through the use of a deformation parameter $q$, yielding the latter measure as $q \to 1$.  The q-deformed Kullback-Liebler measure is 
\begin{equation}
I^{(q)}_{KL}(p,r) = -\int p \ln_{q} {p \over r} \,
\end{equation}
where the deformed logarithm is defined by 
\begin{equation}
\ln_{q} y = {y^{q-1} -1 \over q-1} \, .
\end{equation}
The q-deformed Kullback-Liebler measure reduces to the usual measure as $q \to 1$ while for a reference distribution $r(x)=p(x + L)$ one gets 
\begin{equation}
I^{(q)}_{KL}(p,r) \to -{q L^2 \over 2} I_F(p(x))
\end{equation}
as $L \to 0$. A separable q-deformed multiparticle equation can be obtained by using the action 
\begin{equation}
\Phi_{KL}^{(q)} \equiv {1 \over q} \int dx^{Nd} dt \sum_{i=1}^{Nd} {\cal{E}}_i p(x) \ln_q {p(x) \over p_{i}(x)} \,  \label{qcorrect},
\end{equation}
with $q>0$, instead of eq.(\ref{correct}) in the variational procedure. Note the deformed logarithm, which is actually a rational function as defined above, and the factor $1/q$. The other symbols have the same interpretation as before. In particular, 
\begin{equation}
{\cal{E}}_i L^{2}_{i}= {{\hbar}^2  \over 4 m_{(i)} } \;\;\;\;  \mbox{(no sum on {\it i})} \, . \label{quncert3}
\end{equation}

The nonlinear Schrodinger equation that results is as in (\ref{mpsch}) but with a q-deformed quantum potential
\begin{eqnarray}
Q_{NL}^{(q)} &=& {1 \over q} \sum_{i}^{Nd} {\cal{E}}_i  \left[ \ln_q {p \over p_{+i}} + \left({p \over p_{+i}}\right)^{q-1} - \left({p_{-i} \over p}\right)^q \right] \, . 
\end{eqnarray}
As $q\to 1$ the previous results are recovered. The q-deformation procedure extends obviously to the relativistic case.

Interestingly, the $q$-deformed information measure is also related to some common statistical measures of ``goodness of fit". For example, for  $q=2$ \cite{Kapur}, in discrete variables, $I^{(q)}_{KL}(p,r) = -N \sum (f_e -f_o)^2 /f_o = -N \chi^{2}_{N}$  where $f_e$ are the expected frequencies, $f_o$ the observed frequencies, $N$ the number of data points and $\chi^{2}_{N}$ is Neyman's chi-square function.

\subsection{Regularity and interpolation}
As mentioned, several measures satisfy the properties [C1-C3] listed above, mostly various one or two-parameter deformations of the Kullback-Liebler measure. These will then give rise to classes of nonlinear equations all of which contain a nonzero length parameter with consequent spacetime symmetry breaking. Thus there is no qualitative change in the conclusions of this paper. Those classes of equations  might be useful for some studies, for example to see if the linear Schrodinger equation is in some sense a ``fixed point" in the space of deformed equations. Certainly this is suggested by the fact that the Fisher information is contained in each of them in some limit.   

However all the measures mentioned above, including the Kullback-Libeler mesaure, have one technical problem.  This is the occurence of singularities in the equation of motion, for example when $p(x)$ or $p(x+L)$ vanishes in (\ref{fip}). It is conceivable that in some circumstances such singularities might be physically interesting but in general they are a nuisance. Therefore one other condition that one may  place on a physically useful information measure is:

\begin{itemize}
\item $[C4] \; $ The information measure should not lead to singularities in the equation of motion. 
\end{itemize}

The following measure\footnote{Taken from problem number 6 in Section (7.2.7) of Ref.\cite{Kapur}.} 
\begin{equation}
 -\int p(x) \ln \left[{p(x) \over (1-a) p(x) + a r(x)}\right] \ dx \label{m1}
\end{equation}
with $0 < a \le 1 $ satisfies the additional condition of regularity as we shall see below. For $a = 1$ this measure reduces to the Kullback-Liebler measure.  When used in the variational action with $r(x)=p(x+L)$ it will again lead to a nonlinear Schrodinger equation for which the weak superpostion holds and for which solutions may be scaled. 

However a more useful form for the equations follow if one chooses $r(x)$ to depend on the dimensionless parameter too, $r(x)=p(x+aL)$ as is presently explained.  Consider therefore
\begin{equation}
M^{\eta}_{1} \equiv {1 \over \eta^4 } \int dx^{Nd} dt \sum_{i=1}^{Nd} {\cal{E}}_i p(x) \left[ \ln {p(x) \over (1-\eta) p(x) + \eta p_{i'}(x)} \right] \, \label{M2},
\end{equation}
where now 
\begin{eqnarray}
p_{\pm i'}(x) &=& p(x_1, x_2,.....,x_i \pm \eta L_i,....,x_{Nd}) \,  ,
\end{eqnarray}
and where the dimensionless parameter $\eta$ takes values $0 < \eta \le 1$. This measure gives rise to the generalised potential  
\begin{equation}
Q_{1}^{\eta} = \sum_{i}^{Nd} { {\cal{E}}_i  \over \eta^4}  \left[ \ln {p \over (1-\eta) p + \eta p_{+i'} } + 1 - {(1-\eta) p \over (1-\eta) p + \eta p_{+i'}} - {\eta p_{-i'} \over (1-\eta) p_{-i'} + \eta p} \right] \,  \label{Q2}
\end{equation}
to be used in the separable nonlinear Schrodinger equation (\ref{mpsch}) instead of (\ref{qnon}) and where (\ref{uncert2}) holds. One can now see from the $\eta$-regularised potential that there are no singularities as this would require, for example, that $p$ and $p_{i'}$ vanish simultaneously but in that case the numerator of the relevant term vanishes too.

The nonlinear Schrodinger equation with (\ref{Q2}) will interpolate between the fully nonlinear theory at $\eta =1$ and the usual linear quantum mechanics at $\eta =0$. Notice that at $\eta =0$ one simultaneously recovers the Fisher information measure, the linear theory, and the spacetime symmetries. However even in this limit $L$ is nonzero because the constraint (\ref{uncert2}) is still opertive: $L$ becomes invisible but not zero. 
Unlike $\eta$, the scale $L$ is dimensionful and the limit $L\to 0$  can be performed in some units  only when the constraint (\ref{uncert2}) is adhered to.

One may construct other information measures with the properties [C1-C4] but $M^{\eta}_1$ is probably the simplest, involving a single dimensionless free parameter, $\eta$, that performs the dual role of regulating singularities and allowing interpolation between the linear and nonlinear theories. The parameter(s) $L_i$ with dimension of length appear always, proportional to $\hbar$. The procedure extends obviously to the relativistic case.

\section{Higher derivatives, nonlinearity and nonlocality}

As noted in the examples, the length scale $L$ appears naturally, breaking spacetime symmetries and causing nonlinearities.. Thus within the context of the maximum entropy principle where Kullback-Liebler type information measures are used, nonlinearity cannot be avoided except in some limit where $L$ is scaled appropriately or some other interpolating parameter like  $\eta$ in (\ref{Q2}) is tuned to vanish. When the nonlinearity is eliminated  so is the symmetry breaking. That is, it is only when one uses the Fisher measure as in Ref.\cite{reg1} that an exactly linear quantum theory without Lorentz symmetry breaking is obtained. Thus one would like to understand why the Fisher information measure has such a privileged role in the usual quantum theory. 

As discussed earlier, the variation of the Fisher information measure gives the usual quantum potential in the equations of motion for the hydrodynamic variables $p$ and $S$ that are related to the Schrodinger wavefunction by 
$\psi = \sqrt{p} \ e^{iS/\hbar}$. The structure of the usual quantum potential in turn is completely specified by the structure of the kinetic energy operator in linear quantum mechanics as seen from Eqs.(\ref{sch1},\ref{pot1}). In the coordinate representation the kinetic energy operator ${\hat{p}}^2$ involves second order derivatives as is common to equations of motion in physics. Therefore from this perspective one sees that the special role played by the Fisher information measure is tied to the use of second order derivatives, and no more, for the kinetic terms in the linear Schrodinger's equation. 

Deviations from the above norm can be expected and understood heuristically as follows. If for some reason the spacetime manifold is not smooth at short distances, $L$, then one would expect higher order derivatives in the basic equations, including those of quantum mechanics. Let us assume first that the short-distance Schrodinger equation is linear in the wavefunction, $\psi$, then in terms of the hydrodynamic variables $p,S$ this would result in a quantum potential different from the usual one. In terms of the information theoretic approach this translates into using a different information measure than the Fisher measure. But any information measure other than the Fisher measure implies symmetry breaking {\it and} nonlinearities, as we have seen. 

Another heuristic argument for  symmetry breaking is to note that if in the equation $\hat{p}^2$ is still the kinetic energy operator then higher order derivatives would mean a dramatic change in the generator for translations, $\hat{p} \neq -i\hbar \partial_i$. 

Thus in the qualitative scenario sketched here one expects the Fisher information measure to be only an approximate quantifier that should be replaced by a fuller measure which takes into account higher order derivatives and symmetry breaking, as would be expected at short distances. Nonlinearities in the Schrodinger equation are then a consequence of using the fuller measure.    

It is instructive to view the problem from another perspective, that of the nonlinear equations that have been found in previous sections. Here one has the conventional kinetic energy term but derivatives to all higher orders, with nonlinearities, in the quantum potential. Suppose one could map these equations to a physically  equivalent set described by a linear Schrodinger equation. Then it must be, from the link between the kinetic energy operator and the quantum potential,  that this linear equation contains a highly nonlocal kinetic energy operator that might be approximated by higher order derivatives. Thus it is heuristically conceivable that the nonlinear equations with symmetry breaking that we have found are 
related to linear equations that are highly nonlocal. For practical  applications where one is looking for small deviations and comparing experiments with theory, the nonlinear representation is clearly preferable.  

The above discussion is not totally hypothetical. A concrete example of how a physical system can look very different through a change in variables has already been presented: It is the usual Schrodinger equation. In terms of $\psi$ it is a beautifully linear equation. But in terms of the hydrodynamic variables $p,S$ it is a coupled set of nonlinear differential equations with some nonpolynomial terms like ($1/p$).

\section{Summary of main results}

Using the principle of maximum entropy and thus pushing the arguments of Ref.\cite{reg1} to the extreme, a nonlinear quantum mechanics has been obtained which simultaneously breaks Lorentz symmetry. Both the nonlinearity and the symmetry breaking are determined by a {\it nonzero} length scale $L$ that appears naturally in the definition of the Kullback-Liebler relative entropy (\ref{connection}). The length scale depends on Planck's constant so it is natural to associate $L$  with the Planck length, but it might possibly be a new and larger scale related to the size of elementary particles\footnote{As discussed in Section (4.1), the $L_i$ can be chosen to be spacetime dependent in general, $L_i(x,t)$, suggestive of new dynamical degrees of freedom at short scales.}; some crude estimates suggest that the length might be close to the GUT scale. 

It is intriguing that a  length scale  has emerged without any considerations of gravity or string theory which is where the Planck length traditionally appears \cite{Garay}. Furthermore, although parity and time-reversal symmetries can be preserved even for nonzero length scales within the present formalism, their breaking due to the length scale is just as easily accomodated. 

The phenomena of nonzero length scale, breaking of spacetime symmetries, and nonlinear quantum mechanics are generic outcomes of any relevant information theoretic measure that is used in the inference procedure. {\it Only specially tuned values of some parameters in some measures can give rise to a linear theory and to no symmetry breaking}. Even then, as seen in the example of Section (7.1), the length scale does not vanish but rather becomes hidden because of the relation (\ref{uncert2}).  This  suggests that  the nonlinear equation obtained here is generic and might be relevant for intermediate investigations into the possible simultaneous emergent nature of quantum mechanics and spacetime symmetries.

The nonlinear theory makes some falsifiable predictions. Firstly, since it suggests that the linearity of quantum mechanics and the symmetries of spacetime are intimately linked, the violation of one is expected to lead to the simultaneous violation of the other. Thus, for example, if and when low energy experiments searching for nonlinearity in quantum mechanics do succeed, then experiments searching for Lorentz violation should succeed too in the same energy range. Of course ideally those experiments should be open to both possibilities occuring simultaneously.  

Many more experiments and phenomenological studies have focussed on possible Lorentz and CPT violation \cite{kost,blum}. One much discussed possibility is that certain astrophysical events might be explainable in terms of modified dispersion relations due to the violation or modification of Lorentz invariance \cite{gio}. However the nonlinear Schrodinger equation obtained here still allows the conventional plane waves and also likely contains deformed waves. Thus a second prediction is: At high energies states with the usual dispersion relations should exist and, if the nonlinear equation supports such waves, also those with modified dispersion relations. That is, Lorentz symmetry breaking of the equations of motion does not necessarily imply that all quantum states must have  modified dispersion relations. On the other hand, a mathematically pure plane wave might not exist physically so deformed relations might be more reasonably expected in reality. 

It has been conjectured that the phenomena of neutrino mass and oscillations is due to quantum nonlinearities that are tied to Lorentz violation. Some unusual energy dependence has been noted in this approach and that might be one  way of distinguishing the suggestion of this paper from other proposals. 

There are some issues that deserve further investigation. The nonlinear equation of this paper contains a periodic zero-mode solution which clearly reflects the underlying cellular structure of space. The experimental implications of this are as yet unclear though it might speculatively be relevant in the context of astrophysics and cosmology. More generally, quantum nonlinearities in the equations of motion of particles will obviously impact on the problems of dark matter/energy in cosmology. 

The nonlinear equation (\ref{mpsch}) suggests a natural geometric form of parity and time-reversal violation related to the presence of the length scale $L$. It would be interesting to study the relation of such dicrete symmetry breakings to the conventional discussions. Thus there is the possibility that nonlinear quantum mechanics, CPT breaking and Lorentz violation are all linked.  
 
Finally the effect of the nonlinear terms on the spreading of wavepackets, decoherence of superpositions, and solution of the nonlinear equation with various potentials should be studied \cite{parwani}.

\section{Perspectives}

The construction of physical theories is traditionally driven by various constraints such as: Fundamental lagrangians should not contain third or higher order derivatives; symmetry; renormalizability; gauge invariance and others. Usually there are  also various  implicit assumptions such as the continuity and differentiability of spacetime. Sometimes the constraints are imposed to restrict the possibilities, to simplify the mathematics or to conform with phenomenology. However there is now a growing willingness to explore theories where some of the conventional constraints or assumptions have been relaxed. This is mostly driven either by the desire to understand the origin  of those constraints, or the need to unify quantum mechanics with gravity, or as a probe of potential new physics around the corner.   

We are quite comfortable with Galilean symmetries as they appear reasonable to us, while Lorentzian symmetries are less obvious but were already encoded in Maxwell's equations before Einstein obtained them from more fundamental postulates. 
There have been many suggestions that these symmetries of the continuum might only be approximations that do not hold at smaller length scales. Serious efforts to search for violations of Lorentz invariance are under way. On the other hand, the linearity of quantum mechanics has been a long standing puzzle that has engaged many authors, with several suggestions that it  might be an approximation of a nonlinear theory. Experiments that have been performed to search for departures from linearity have to date concluded that the deformation parameters must be tiny if not zero.

If, for example, quantum effects lead to fluctuations of the geometry of spacetime at small scales, and perhaps even render the notion of space and time meaningless, then one needs a guiding principle other than the usual one based on symmetries that has sucessfully led us so far to the Standard Model of particle physics.  Wheeler has forcefully expressed his ``It from Bits" view that an information theoretic approach is the way to go in uncovering the basic foundation of our laws \cite{Wheeler}. Jaynes \cite{jay} and others have shown how an inference method can be used successfully in statistical mechanics and many other fields. Frieden and Reginatto used a similar approach in obtaining the usual Schrodinger equation.  In this paper those initiatives have been pushed further to see what new physics might lie below the surface. A surprising but pleasing conclusion has been reached, linking two apparently unconnected aspects of physics: The linearity of quantum mechanics and our accepted spacetime symmetries.

The excellent spacetime symmetries and the superb linearity of quantum mechanics appear in current physics as two separate pieces of fact. It is proposed here that those facts are actually intrinsically linked: Both are approximations controlled by the same deformation parameter, a nonzero length scale. The main argument has relied on the assumption that a common information-theoretic principle be used to infer the dynamics of quantum theory --- the information measure should be similar to that used in the maximum entropy principle. Other supporting and suggestive arguments linking the linearity of quantum mechanics with causality have been noted in Section (4.2). 

Of course one can construct a nonlinear Schrodinger equation which respects spacetime symmetries or a linear equation that breaks those symmetries, as many authors have done within various other  contexts. One can also construct any number of equations that are simultaneously nonlinear and break spacetime symmetries. What has been done here is to use the
framework of information theory, that has been successfully used elsewhere, to restrict the possible forms a generalised Schrodinger equation can take, resulting in a link between two seemingly different aspects of conventional physics.

The conclusion reached here gives prominence to information theoretic arguments and has ignored any initial symmetry constraints.  Alternatively if one insisted on the fundamental validity of current spacetime symmetries and imposed them at the beginning, the results obtained here could be reinterpreted as : ``Information theoretic arguments supplemented with spacetime symmetry constraints imply the linearity of quantum mechanics". Thus the observed linearity of quantum mechanics would be a consequence of ``hidden symmetries" --- hidden not because  they were  not manifest in the equations but because they were not previously seen to be operational in determining the linearity! Whichever perspective one adopts, that of the previous paragraph or this, the main point is the linking of spacetime symmetries with the linearity of quantum mechanics. 

However the imposition of symmetry constraints at the outset within the information theoretic approach would allow the use of only the Fisher information measure. As discussed in Section (8) this can be understood as due to certain other assumptions that are usually made about the structure of quantum mechanics, with the conclusion that linearity and symmetries would anyway be expected to fail at short distance scales.   

Ultimately these arguments have to confront experiments, and some tentative predictions and suggestions were made in Section(9). For example, current searches for violations of Lorentz invariance might also be unwittingly searching for violations of the linearity of quantum theory. If data is limited, it is conceivably possible to attribute it wholly to violations of Lorentz invariance or wholly to violations of linearity of quantum mechanics, so a more elaborate analysis would be required to see if both violations are indeed occuring simultaneously. The unusual energy dependence found here might be one signature.

Certainly it is not being suggested that the nonlinear equation obtained in this paper is the fundamental equation underlying quantum matter and spacetime but rather, as discussed in Section (8),  it might be one approximation in a sequence leading to a final theory (if a unique theory exists!) underlying space, time and matter. The equation may take different guises in different variables. If one insists on using the wavefunction $\psi$ then nonlinearities and nonlocalities are to  be expected at short distances. However at a more basic level there might be different variables in terms of which the  equations take a more pleasing form. A final theory might involve radically different concepts, as discussed for example in Ref. \cite{final}, which at a coarser level give rise to our perception of  linear quantum matter and smooth spacetime.  

In some sense, the information theoretic method used here to  uncover a more basic structure to our current theories may be viewed as an inverse procedure  to the Wilson renormalisation scheme where one obtains successive effective low energy theories  by integrating out the unobserved high energy modes. In the inference method discussed here, one is ``integrating back in" the fluctuations. Can an inference procedure also be iterated to obtain successively better theories? 

\section*{Appendix: Fields and strings}

Formally the inference procedure described for nonrelativistic quantum mechanics can be applied to quantum field theory. One works in the Schrodinger representation and writes the functional Schrodinger equation \cite{holland,hatfield} for the wave-functional $\Psi[\phi(x),t]$ which describes the state of time-independent fields $\phi(x)$. The usual transformation $\Psi = R e^{iS/\hbar}$ produces the continuity equation and a quantum Hamilton-Jacobi equation. The quantum potential, which is of Fisher form, can then be generalised as usual leading finally to a nonlinear functional Schrodinger equation.

Apart from the usual possible ambiquities with functional equations, the main problems are relating the generalised results to the single particle theories considered so far and extracting some physical consequences \cite{parwani}.  

It is possible that the procedure just described for fields will reduce to different single particle theories from that described in the main text. However this is not surprising as one is not adhering to the conventional quantisation approach.
Indeed, one might {\it define} a nonlinear quantum field  theory in several possibly inequivalent ways. For example, the nonlinear Klien-Gordon equation of the text can be promoted to an operator equation (Heisenberg representation). Or instead the $\phi$ can be treated as a field and the nonlinear Lagrangian then used directly in a path-integral.  The path-integral {\it ansatz} probably retains the closest link with the nonlinear single particle equations considered so far.

It might be interesting to study also the nonlinearisation of quantum bosonic strings \cite{hatfield}. The obvious question is whether the  tachyonic state of the linear theory persists.

\end{document}